\documentclass[%
 reprint,
superscriptaddress,
 amsmath,amssymb,
 aps,
]{revtex4-2}
\usepackage{ulem} 
\usepackage{graphicx}
\usepackage{dcolumn}
\usepackage{bm}
\usepackage{xcolor}
\usepackage{gensymb}
\usepackage[detect-all]{siunitx}
\usepackage[utf8]{inputenc}
\usepackage{textcomp}

\begin{document}
\preprint{APS/123-QED}
\title{Doping dependence of the low temperature planar carrier density in overdoped YBa$_2$Cu$_3$O$_{7-\delta}$}

\author{R. Nicholls} 
\email{beckie.nicholls@bristol.ac.uk}
\affiliation{These authors contributed equally to this work.}
\affiliation{H. H. Wills Physics Laboratory, University of Bristol, Bristol BS8 1TL, United Kingdom}

\author{R. D. H. Hinlopen} 
\email{roemer.hinlopen@mpsd.mpg.de} 
\affiliation{These authors contributed equally to this work.}
\affiliation{H. H. Wills Physics Laboratory, University of Bristol, Bristol BS8 1TL, United Kingdom}

\author{J. Ayres} 
\affiliation{H. H. Wills Physics Laboratory, University of Bristol, Bristol BS8 1TL, United Kingdom}

\author{T. Kotte} 
\affiliation{Dresden High Magnetic Field Laboratory (HLD-EMFL), Helmholtz-Zentrum Dresden-Rossendorf (HZDR), Dresden, Germany}

\author{T. F\"orster} 
\affiliation{Dresden High Magnetic Field Laboratory (HLD-EMFL), Helmholtz-Zentrum Dresden-Rossendorf (HZDR), Dresden, Germany}

\author{J. Park} 
\affiliation{Dresden High Magnetic Field Laboratory (HLD-EMFL), Helmholtz-Zentrum Dresden-Rossendorf (HZDR), Dresden, Germany}

\author{J. Sourd} 
\affiliation{Dresden High Magnetic Field Laboratory (HLD-EMFL), Helmholtz-Zentrum Dresden-Rossendorf (HZDR), Dresden, Germany}

\author{A. Carrington}
\affiliation{H. H. Wills Physics Laboratory, University of Bristol, Bristol BS8 1TL, United Kingdom}

\author{N. E. Hussey}
\email{n.e.hussey@bristol.ac.uk}
\affiliation{H. H. Wills Physics Laboratory, University of Bristol, Bristol BS8 1TL, United Kingdom}
\affiliation{High Field Magnet Laboratory (HFML-FELIX) and Institute for Molecules and Materials, Radboud University, Nijmegen 6525ED, Netherlands}

\date{\today}

\begin{abstract}

Whether a quantum critical point (QCP) demarcates the end of the pseudogap (PG) regime in hole-doped cuprates at a singular doping level $p^*\approx0.19$ is still an open question. A crucial component of the puzzle is the manner in which the carrier density predicted by electronic structure calculations is recovered for $p>p^*$. Here, we use magnetic fields up to 67~T to suppress superconductivity down to 50~K, allowing us to simultaneously measure the low temperature Hall number $n_\textrm{H}$ and the in-plane resistivity anisotropy $\rho_a/\rho_b$ in overdoped Y$_{1-x}$Ca$_x$Ba$_2$Cu$_3$O$_{7-\delta}$ single crystals. We confirm a previous finding [Badoux \textit{et al.}, \textit{Nature} \textbf{531}, 210 (2016)] that $n_\textrm{H}$(50~K) exhibits a sharp increase below $p^{\ast}$ and use the measured resistivity anisotropy to obtain the planar carrier density $n_{pl}=n_\textrm{H}(\rho_{a}/\rho_{b})^{-1}$. The doping dependence of $n_{pl}$(50~K) reveals two key findings: (i) at optimal doping, $n_{pl}\approx p$ and (ii) the sharp rise in $n_{\rm H}(p)$ is softened to such an extent that the full Fermi volume ($n_{pl}=1+p$) is only partially recovered at $p^{\ast}$. The latter result disfavors a conventional QCP scenario in which the PG end point leads to a reconstructed Fermi surface.

\end{abstract}


\maketitle

One of the most striking features of the cuprate phase diagram is the emergence of a near-parabolic dome of high-temperature superconductivity out of a Mott insulating parent state \cite{Keimer2015}. As the parent crystal is doped with $p$ holes per Cu atom, the superconducting transition temperature $T_c$ rises to a maximum at optimal doping ($p_{opt}=0.16$) before falling back to zero at the edge of the superconducting (SC) dome ($p_{sc}\approx0.3$). For $p > p_{sc}$, Fermi liquid (FL) behavior is recovered \cite{Proust2002, Nakamae2003} and the Fermi volume contains $1+p$ holes \cite{Rourke2010}. At lower dopings, however, a reduction in the carrier density suggests a mixture of itinerant and localized holes. The exact doping level at which the hole initially localized in the Mott state becomes fully itinerant is still an open question, and the mechanism(s) of interaction between the itinerant and localized holes are still unclear. The Hall effect can provide information about the planar carrier density $n_{pl}$ and its evolution with doping across the relevant part of the phase diagram. However, the very high magnetic fields needed to suppress the SC state restrict measurements of the low-$T$ Hall coefficient $R_\textrm{H}=1/(en_\textrm{H})$ while the assumed relationship $n_\textrm{H}=n_{pl}$ is complicated by the presence of the pseudogap \cite{Carrington1992}, van Hove singularities \cite{Narduzzo2008}, complex ordering phenomena \cite{Balakirev2003, LeBoeuf2007} as well as possible incoherent contributions to the conductivity \cite{Putzke2021}. 

In a seminal study, Badoux \textit{et al.}~measured $R_\textrm{H}$ in optimally doped and overdoped YBa$_2$Cu$_3$O$_{7-\delta}$ (Y123) in fields of up to 88 T \cite{Badoux2016} and found a rapid rise in the Hall number from $n_{\rm H}(50\textrm{~K}) \approx p$ at optimal doping to $n_{\rm H}(50\textrm{~K})$ $\approx 1 + p$ at $p = 0.205$. This sharp, six-fold increase in $n_{\rm H}$ has been interpreted as evidence for a quantum phase transition associated with the closing of the PG at $p^\ast=0.19$ \cite{Storey2016, Su2024}. Subsequent high-field Hall \cite{Collignon2017} and specific heat \cite{Michon2019} measurements on Nd-doped La$_{2-x}$Sr$_x$CuO$_4$ (Nd-LSCO) reached a similar conclusion, albeit with a higher value of $p^*=0.23$.

Since these original studies, however, conflicting evidence has emerged of a far more gradual recovery of the localized hole upon doping. High-field transport studies of Tl$_2$Ba$_2$CuO$_{6+\delta}$ (Tl2201) and Pb/La-doped Bi$_2$Sr$_2$CuO$_{6+\delta}$ (Bi2201) \cite{Putzke2021} as well as time-domain spectroscopy studies of LSCO \cite{Legros2022} find that the $p$ to $1+p$ crossover occurs over a much broader doping range between $p_{opt}$ and $p_{sc}$, with no notable feature coinciding with $p^{\ast}$. Over the same doping range, the in-plane resistivity $\rho_{ab}(T)$ contains a non-FL $T$-linear component down to low-$T$ whose coefficient $\alpha_1$ scales with $T_c$ \cite{Cooper2009, Hussey2013, Putzke2021}. This non-FL form of $\rho_{ab}(T)$ shows that the FL state is not immediately recovered at $p^*$ as one might expect in a conventional QCP scenario when traversing the $T=0$ line  \cite{Hussey2018}. Similarly, the in-plane magnetoresistance (MR) exhibits anomalous $H$-linear behavior at high fields \cite{Giraldo-Gallo2018, Ayres2021} with a $T$-independent slope that also scales with $\alpha_1$ and $T_c$ \cite{Ayres2024}. These collective phenomena signify an extended non-FL regime that spans the majority of the overdoped side of the SC dome \cite{Hussey2018}.

A pertinent question then arises -- how to reconcile the very different character of the $p$ to $1 + p$ crossover in Y123 and Nd-LSCO with those found in other cuprates? With regard to the former, a simple but perhaps crucial factor that was acknowledged, but not explored in the initial study \cite{Badoux2016}, is the role of quasi-one-dimensional (quasi-1D) conducting CuO chains oriented along the $b$-axis. The parallel resistor model in the Supplemental Material \cite{SupplementalMaterial} (see also \cite{Segawa2004,Ong1991}) shows how the quasi-1D current carried by these chains modifies the relation between $R_\textrm{H}$ and $n_{pl}$ such that $n_{pl}=n_\textrm{H}(\rho_{a}/\rho_{b})^{-1}$. Above $T_c$, $\rho_{a}/\rho_{b}$ is strongly dependent on both temperature and the degree of oxygenation within the chains \cite{Ando2002}. However, this anisotropy ratio has not been systematically measured in the relevant low-$T$, high-$H$ limit. 

Here, we present pulsed field measurements of the $T=50$~K resistivity anisotropy and Hall effect in bulk Y123 crystals in the overdoped regime (0.16 $\leq p \leq$ 0.19). By measuring both quantities for each detwinned single crystal, we eliminate any disorder- or doping-induced discrepancies. First and foremost, we find that our $R_{\rm H}$ data are consistent with previous reports \cite{Badoux2016,Segawa2004} and, after taking into account the resistivity anisotropy, reaffirm the surprising result that at optimal doping, $n_{pl} \approx p_{opt} < n_s$ -- the superfluid density. Crucially, we uncover a \textit{gradual} rise in $n_{pl}$ with increasing doping in Y123, with $n_{pl}$ remaining well below $1+p$ at $p^*$ and extrapolating to $n_{pl}=1+p$ only towards the end of the SC dome. In this way, the doping dependence of the Hall number in Y123 appears to be reconciled with previous findings in Bi2201, Tl2201 and LSCO \cite{Putzke2021, Legros2022, Culo2021_cuprates}. 

\begin{figure}[b]
\centering
	\includegraphics[width=0.48\textwidth]{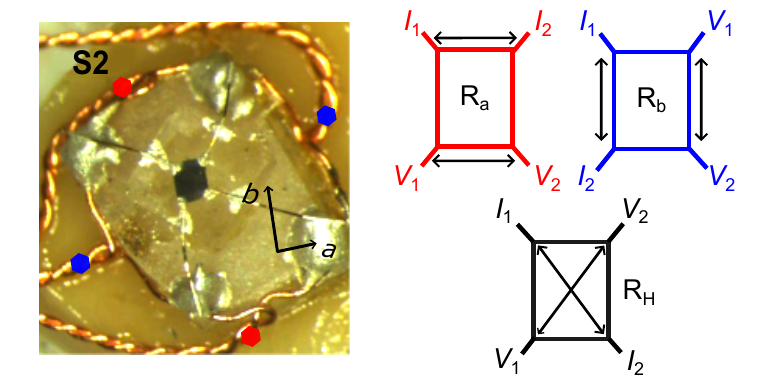}
	\caption{Detwinned rectangular Y123 single crystal (sample S2, $p\approx 0.17$) mounted on a quartz platelet using 25~$\mu$m gold wires and conductive silver paint. The in-plane crystallographic directions are also labelled. Contacts wrap around the sides of the sample to short out the $c$-axis. Four twisted pairs of Cu wires allow the three $IV$ combinations to be measured without altering the connections on the quartz mount. }
	\label{fig:Montgomery}
\end{figure}

\begin{table}[t]
\begin{ruledtabular}
\begin{tabular}{ccccc}
Sample & Ca (\%) &   $T_c$ (K)   & $\Delta T_\textrm{c}$ (K) & $p$   \\
\hline
S1   & 0 & 93.8$^*$  & 0.15   &    $0.160\pm0.003$    \\
S2 & 5 & 93.1$^*$  & 0.12      &     $0.170\pm0.002$ \\
S3 & 2 & 93.1$^*$ & 0.10     &       $0.170\pm0.002$      \\
S4 & 15 & 92.0$^\dag$ & 0.7      &  $0.175\pm0.002$   \\
S5 & 15 & 91.0$^\dag$ & 0.8      &  $0.179\pm0.002$   \\         
S6 & 15 & 87.2$^\dag$ & 1.1    &   $0.189\pm0.002$  \\  
S7 & 15 & 86.2$^\dag$ & 1.3    &   $0.191\pm0.002$  \\        
\end{tabular}%
\caption{\label{tab:Samples} List of Y123 samples investigated in this study. The Ca percentage is derived from the nominal (Ca+Y)/Ca fraction in the precursor materials. EDX measurements of crystals from various nominal 15\% growths showed actual Ca \% between $5-8$\%. $T_c$ is taken as the midpoint of the zero-field resistive transition; symbols denote annealing conditions of either ($*$) 1~bar O$_2$ at 500~$^\circ$C or ($\dag$) $\sim30$ bar O$_2$ at 450~$^\circ$C. $\Delta T_c$ is the 10–90\% superconducting transition width.  Hole doping $p$ is calculated using the empirical relationship in Ref.~\cite{Liang2006} with $T_c^{\rm max}$ = 93.8~K. }
\end{ruledtabular}
\end{table}

High quality single crystals of Ca-free and Ca-doped Y123 were grown via the self-flux method using yttria-stabilized zirconia (YSZ) crucibles and high purity Y$_2$O$_3$, BaCO$_3$, CaCO$_3$ and CuO \cite{Liang2012}. Substitution of a fraction of the Y$^{3+}$ cations with Ca$^{2+}$ is necessary for hole-doping beyond stoichiometric Y123 ($\delta=0$, $p\approx 0.18$). The as-grown crystals (nominal $x=0-0.15$) were annealed for several days in either a continuous flow of high purity oxygen or in a sealed high-pressure oxygen environment, as detailed in Table~\ref{tab:Samples}. Crystals were detwinned by heating to $\sim$ 250$^\circ$C and applying uniaxial stress until twin lines were no longer visible under polarized light.

The current-voltage ($IV$) configurations used to measure the in-plane resistances ($R_a$, $R_b$) and Hall coefficient ($R_{\rm H}$) are shown in Figure~\ref{fig:Montgomery}. Pulsed field measurements were conducted at HLD-EMFL at fixed temperatures ($T<T_c$) with maximum field strengths ranging from $57-67$~T. The in-plane resistivities were extracted by the Montgomery method \cite{Montgomery1971,DosSantos2011} and the Hall signal was decoupled from any longitudinal components by anti-symmetrizing the voltages measured across the diagonals. A more thorough description of the experimental procedure is given in the Supplemental Material \cite{SupplementalMaterial}. Measurements of $\rho_a$, $\rho_b$ and $R_\textrm{H}$ at $T$ = 50~K were made for all samples except S7 for which $R_\textrm{H}(50{\textrm{~K}})$ could not be measured due to contact failure. However, as our $R_{\textrm{H}}$ data match those in Ref.~\cite{Badoux2016} where it is shown that $R_{\rm H}(50\textrm{~K})\approx R_\textrm{H}(80\textrm{~K})$ for $p=0.19$, static field measurements of $R_\textrm{H}(80\textrm{~K})$ ($\mu_0 H$ = 14~T) have been used in the analysis of S7. Note that these data points are differentiated by color in subsequent figures.

\begin{figure*}[t]
\centering
	\includegraphics[width=0.98\textwidth]{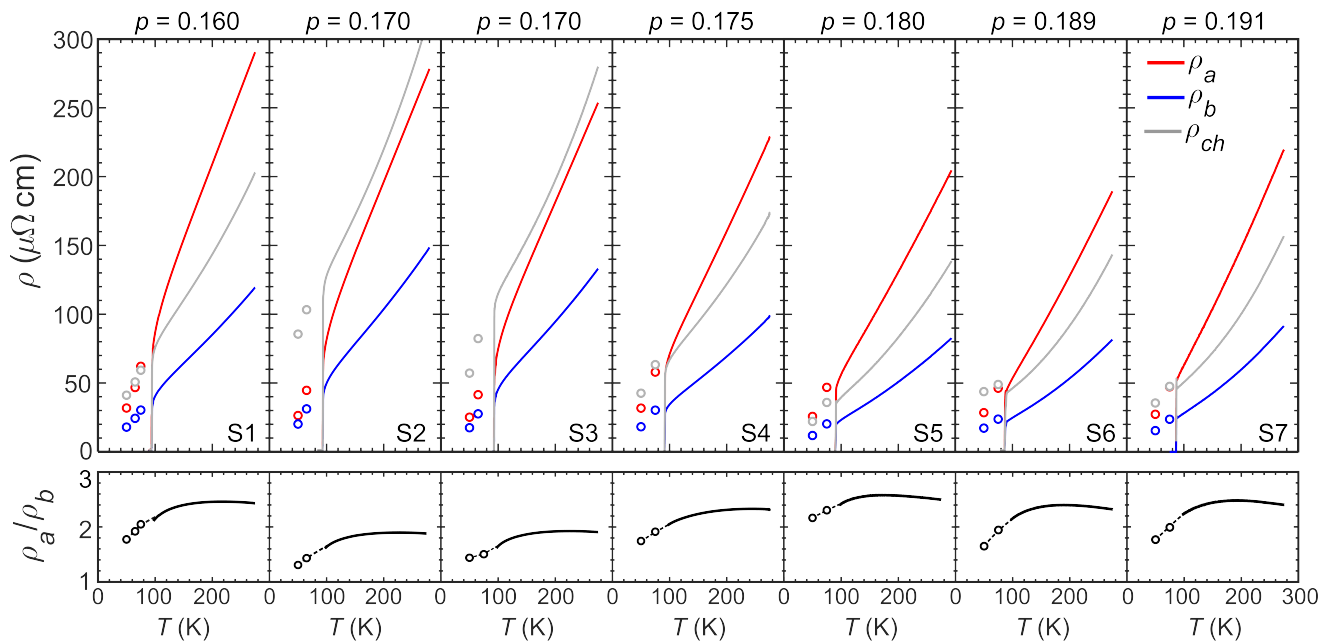}
	\caption{(Top panels) Temperature dependence of the in-plane $\rho_a(T)$, $\rho_b(T)$ and chain $\rho_\textrm{ch}(T)$ resistivities for seven Y123 crystals with hole dopings between $p=p_{opt}$ and $p\approx p^\ast$. Solid lines were obtained from Montgomery analysis of zero-field resistance measurements using an excitation current of 1 mA. Open circles are from Montgomery analysis of data recorded at field strengths of 57-67~T using an excitation current of 10~mA, see Figure~S2 in the Supplemental Material \cite{SupplementalMaterial}. (Bottom Panels) Resistivity anisotropy ratios for the respective samples. Dashed lines are guides to the eye.} 

	\label{fig:ResistivityAnisotropies}
\end{figure*}

Figure~\ref{fig:ResistivityAnisotropies} shows $\rho_{a}(T)$, $\rho_b(T)$ and the associated anisotropy ratios for the seven Y123 crystals studied. The magnitude of the anisotropy ratio varies between samples,  but all $\rho_{a}/\rho_b(T)$ curves are similar to those reported for Ca-free Y123 with nearly-filled chains ($\delta=0-0.17 $) \cite{Ando2002}. Also shown in Figure~\ref{fig:ResistivityAnisotropies} (faint gray lines) is the chain resistivity $\rho_\textrm{ch}(T)$ calculated assuming a parallel resistor model ($\rho_\textrm{ch}^{-1}=\rho_\textrm{b}^{-1}-\rho_\textrm{a}^{-1}$), which serves as a measure of the defect density in the chains.
 
The $\rho_a(T)$ curves near $p_{opt}$ exhibit the well-known $T$-linear response down to a characteristic temperature $T^\ast \sim$ 160-190 K associated with the pseudogap \cite{Ito1993}, though linear fits to these curves have a negative intercept at $T = 0$. Such negative $\rho_{a}(0)$ values have been seen before in optimally doped Y123 \cite{Yoshida1999} and imply that the $T$-linear resistivity above $T^\ast$ must cross over to a higher exponent power-law dependence at low $T$. Indeed, exposing the normal state resistivity below $T_c$ using high magnetic fields has previously revealed pure $T^2$ behavior of the in-plane resistivity in a host of underdoped cuprates, including Y123 \cite{Rullier2007, Chan2014, Proust2016, Berben2022}.

\begin{figure}[t]
\centering
	\includegraphics[width=0.4\textwidth]{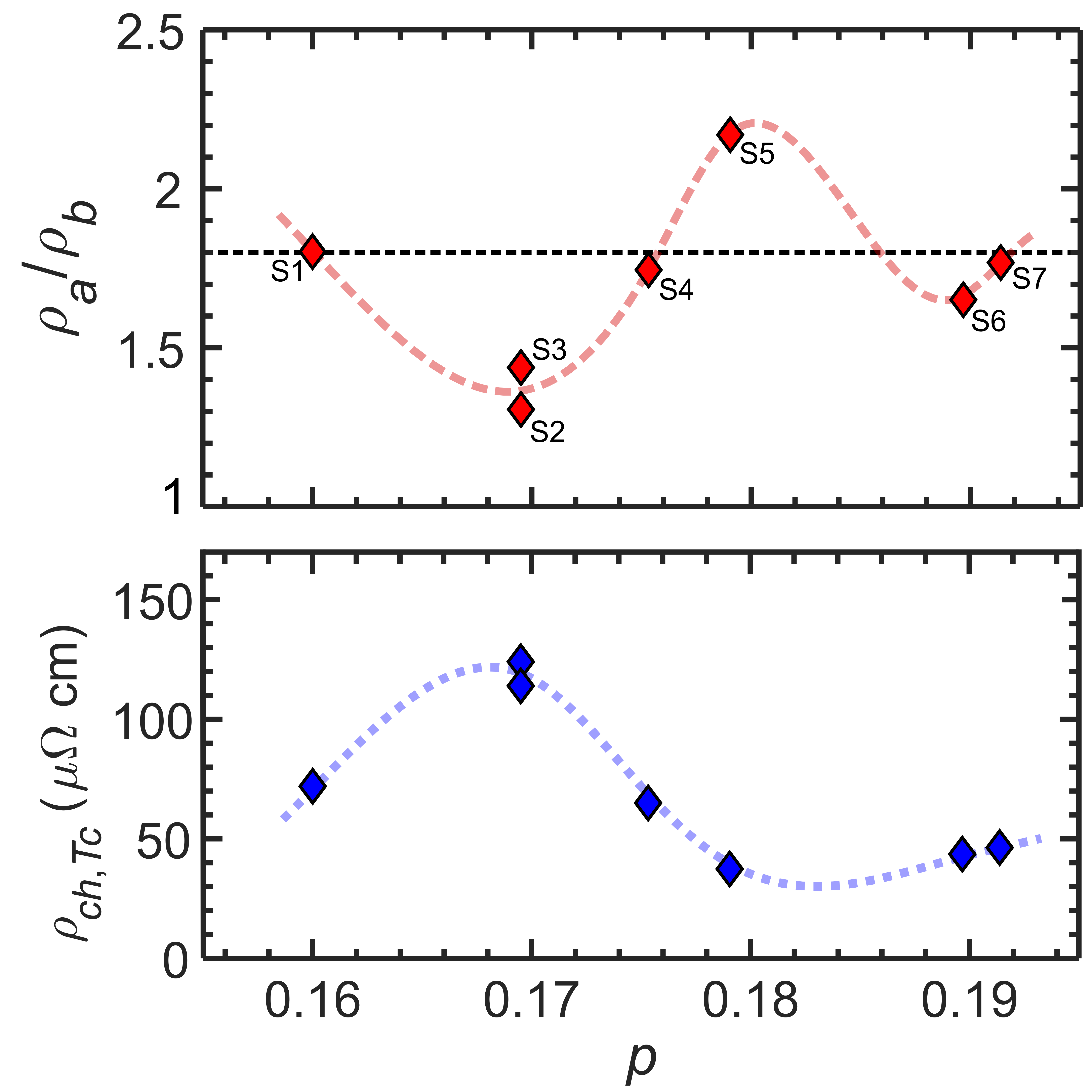}
	\caption{(Top) Resistivity anisotropy ratios for Y123 crystals at $T = 50$~K. The horizontal black dashed line indicates the average ratio for the four samples with the most conductive chains, $\rho_a/\rho_b \approx 1.8$. (Bottom) Chain resistivity $\rho_{ch}$ near $T_c$. The colored dashed lines are guides to the eye.}

	\label{fig:ChainInfluence}
\end{figure}

The top and bottom panels of Figure~\ref{fig:ChainInfluence} show the doping dependence of $\rho_a/\rho_b$(50~K) and $\rho_{ch}(T\approx T_c)$, respectively. The anti-correlation between the two quantities is clear.  Specifically, the optimally doped Ca-free crystal ($p=0.16$, $\delta=0.06$) and the four Ca-doped samples annealed under high pressure ($p>0.17$, $\delta\approx 0$) exhibit the lowest chain resistivities and the strongest resistivity anisotropies, averaging $\rho_a/\rho_b=1.8\pm0.2$. The anisotropy ratio of the two Ca-doped samples annealed in 1 bar O$_2$ ($p=0.170$) is lower ($\approx1.35$), consistent with previous reports of reduced oxygen uptake in Ca-doped Y123 crystals \cite{Chandrachood1990, Titova2016, Yakabe_1995, Fisher1993}. This sensitivity of $\rho_a/\rho_b$ to the degree of chain disorder highlights the need to measure the ratio independently on each crystal when extracting $n_{pl}$ from $n_{\rm H}$ in Y123.

\begin{figure}[t]
\centering
	\includegraphics[width=0.48\textwidth]{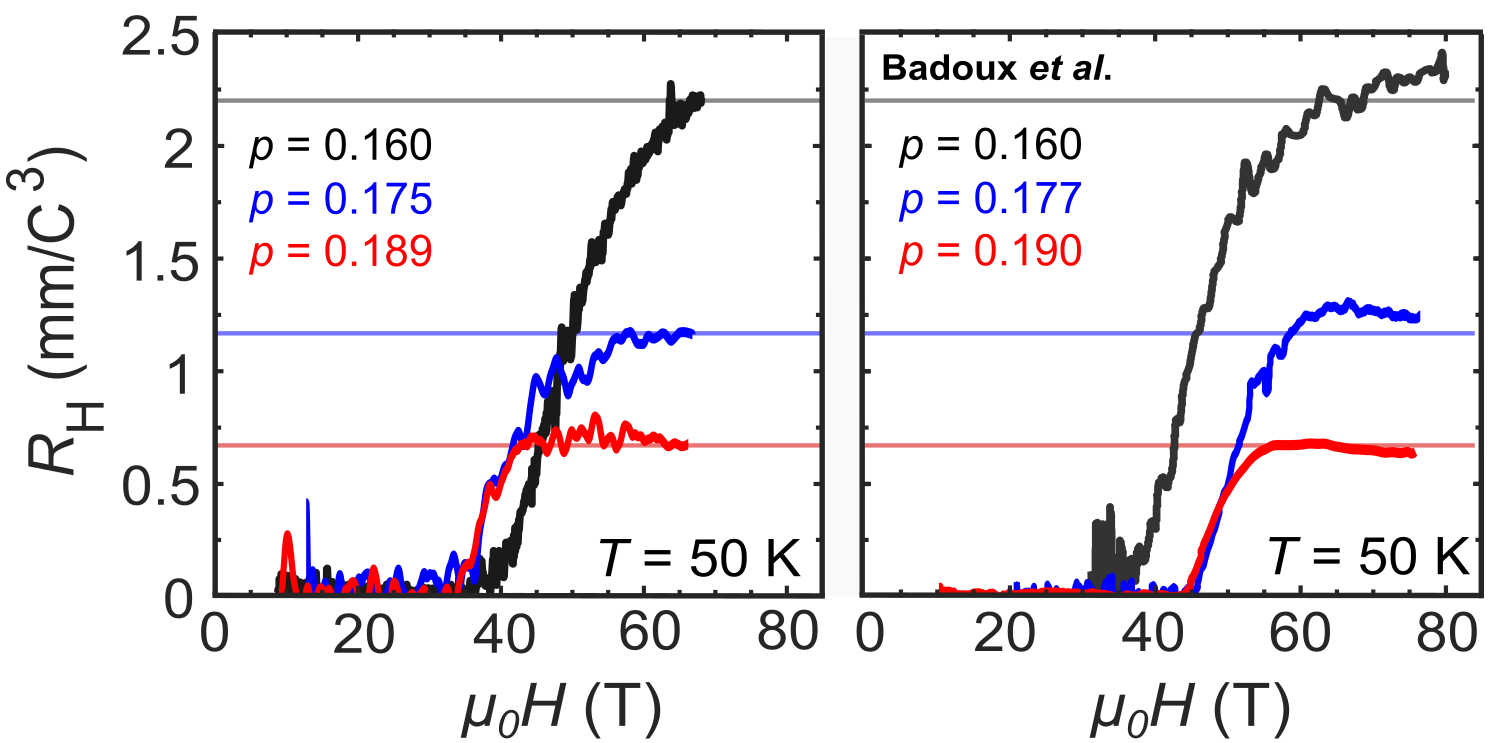}
	\caption{(Left) Pulsed field data up to 67~T, this work. Faint solid lines extending into the right-hand panel mark $R_\textrm{H}$(65~T) for each new curve. (Right) Data extracted for comparable crystals from Ref.~\cite{Badoux2016}. Data shown only for $H>10$~T due to the divergence of $R_\textrm{H}=R_{xy}/\mu_0H$ as $H\to0$. }
	\label{fig:HallComp}
\end{figure}

 \begin{figure*}[t]
\centering
	\includegraphics[width=.98\textwidth]{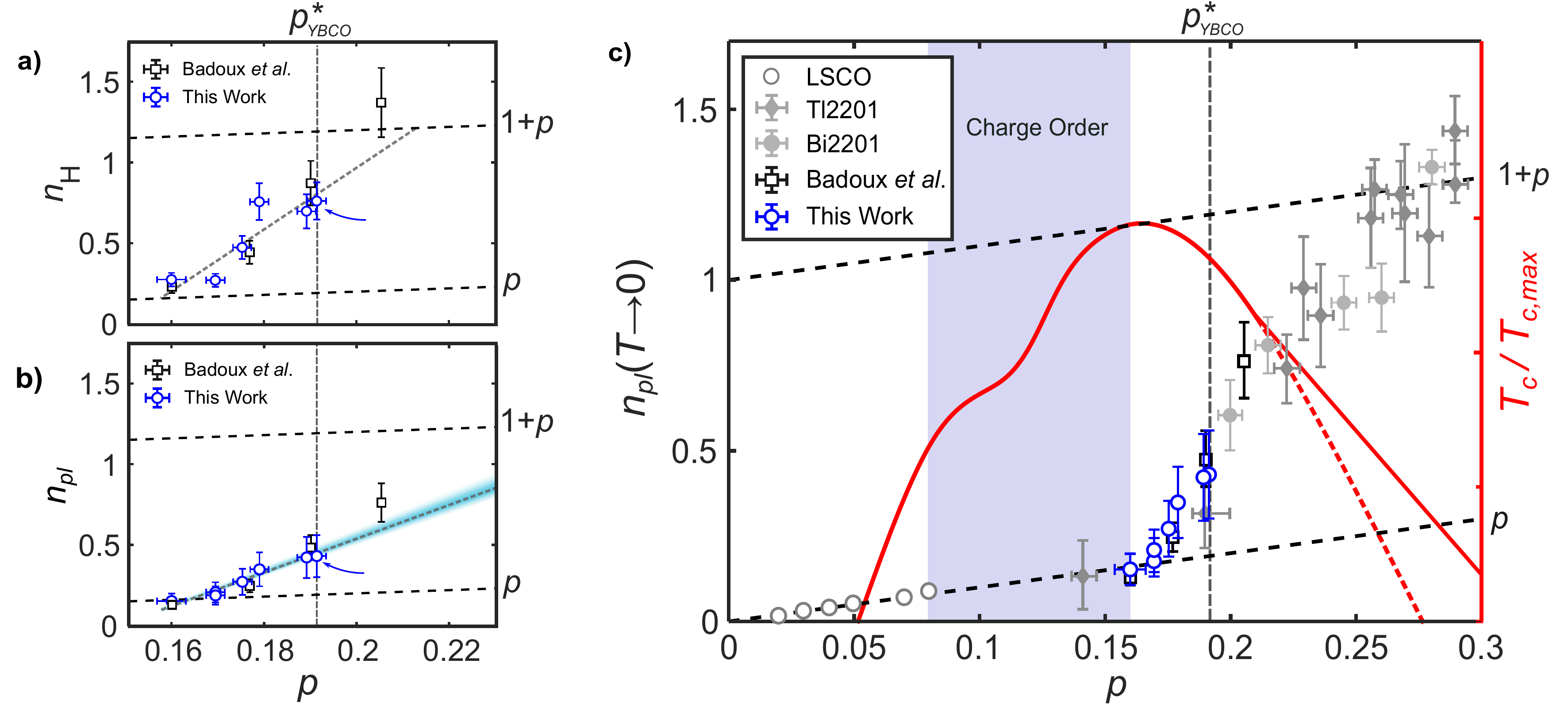}
	\caption{\textbf{a)} Doping evolution of the measured Hall carrier density in Y123 at $T = 50$~K. Black squares are data extracted from Ref.~\cite{Badoux2016}. Blue circles show $n_\textrm{H}$(50~K) measured in this study, aside from the data point marked with a blue arrow (S7, $p=0.191$) where $n_\textrm{H}$(80~K) -- see main text for details. Black dashed lines represent $n_{\rm H} = p$ (the number of doped holes) and $n_{\rm H} = 1 + p$ (the full Fermi volume). Error bars of $\pm15$~\% reflect uncertainty in the dimensions of the crystals, the finite contact areas and contact misalignment. Gray dashed line is linear fit from panel (b), scaled up by a factor 1.8. \textbf{b)}  Doping evolution of the planar carrier density in Y123, $n_{pl}=n_{\rm H}(\rho_a/\rho_b)^{-1}$.  Data from this work use as-measured values of $\rho_a/\rho_b$(50~K). Error bars of $\pm30$~\% reflect the additional geometric uncertainties involved in the anisotropy term. Data from Ref.~\cite{Badoux2016} have been rescaled by a generic anisotropy factor $\rho_a/\rho_b = 1.8$, as discussed in main text. Gray dashed line is a weighted linear fit to all points.  Blue shaded region shows the range of linear fits to \textit{all} data points when the four $n_{\textrm{H}}$ values from Ref.~\cite{Badoux2016} are rescaled by a singular anisotropy ratio in the range $1.5 \leq \rho_a/\rho_b \leq 2.1$. This does not represent a definitive boundary. \textbf{c)} Doping evolution of the low-$T$ planar carrier density in Y123, Bi2201 and Tl2201. The Y123 data ($n_{pl}$) are reproduced from panel (b), while the Bi2201 and Tl2201 data are reproduced from Ref.~\cite{Putzke2021}. Dashed red line indicates the proposed form and extent of the SC dome in LSCO and Bi2201 at high dopings \cite{Berben2022}, while the solid red line relates to Tl2201, where the dome extends to $p\approx0.31$ \cite{Putzke2021}.}
	\label{fig:nH_All}
\end{figure*}

In Figure~\ref{fig:HallComp}, we compare our raw $R_{\rm H}(H)$ curves to those reported in Ref.~\cite{Badoux2016} for samples with equivalent dopings. While the irreversibility fields differ between comparable samples, our high-field values of $R_\textrm{H}$ and the functional forms of $R_\textrm{H}(H)$ are in good agreement with the original study. Figures~S3 and S4 in the Supplemental Material \cite{SupplementalMaterial} show the remaining $R_{\rm H}(H)$ curves and the high-$T$ Hall data for S4, S5 and S7, respectively. The measured Hall carrier density is extracted from $n_\textrm{H}=V_\textrm{cell}/(2eR_\textrm{H})$, where $V_\textrm{cell}$ is the unit cell volume. As changes in the lattice parameter with $T$, O and Ca content are less than $2$~\% within the measured regimes, a constant value of $V_\textrm{cell}=174$~{\AA}$^3$ \cite{Jirak1987} has been used throughout.

Figure~\ref{fig:nH_All}a shows $n_\textrm{H}$(50~K) for S1-6 (red) and $n_\textrm{H}$(80~K) for S7 (black) alongside the data from Ref.~\cite{Badoux2016} (blue). It is interesting to observe that the deviation away from a linear dependence on $p$ mirrors the variation in $\rho_a/\rho_b$ across the series, foreshadowing the impact of the resistivity anisotropy on $n_\textrm{H}$. In Figure~\ref{fig:nH_All}b, we plot the corresponding planar carrier densities $n_{pl}=n_{\rm H}(\rho_a/\rho_b)^{-1}$ to re-examine the extent of the $p$ to $1+p$ crossover. For completeness, the $n_{\rm H}$ data from Ref.~\cite{Badoux2016} have been included and renormalised by $\rho_a/\rho_b = 1.8$ (horizontal dashed line in Fig.~\ref{fig:ChainInfluence}, representative of samples with nearly full chains). The blue band in Figure~\ref{fig:nH_All}b shows that if the data in Ref.~\cite{Badoux2016} are rescaled by a range of $\rho_a/\rho_b$ values and included in the overall linear fit, the $n_{pl}$ fit remains far below the 1+$p$ line as $p \rightarrow p^*$. Figure~S5 in the Supplemental Material \cite{SupplementalMaterial} shows the equivalent plot using only data from Ref.~\cite{Badoux2016}.

Figure~\ref{fig:nH_All}c shows the resultant comparison of $n_{pl}(T\to 0)$ in Y123, Tl2201 and Bi2201, where the combined $n_{pl}$ values trace out a quasi-linear increase from $n_{pl}=p$ at the end of the charge-ordered regime to $n_{pl}=1+p$ at the edge of the SC dome. The coincidence of the three data sets suggests a universal mechanism governing the emergence of mobile carriers across $p^\ast$ and into the overdoped \lq strange metal' regime. 

The one outlier to this trend is Nd-LSCO (not shown) which exhibits a strong increase in $n_{\rm H}$ above $p = 0.20$, with $n_{\rm H}(0) \approx 1 + p$ at $p \sim p^{\ast} = 0.23$ \cite{Collignon2017}. Notably, this doping level lies close to the Lifshitz transition point at which the FS develops both electron-like and hole-like character. Such duality in the FS curvature leads to contributions to the Hall conductivity $\sigma_{xy}$ of opposite sign, thereby reducing $R_{\rm H}$ and increasing the effective $n_{\rm H}$. Hence, it is likely that a large part of the increase in $n_{\rm H}$ in Nd-LSCO is in fact due to the Lifshitz transition rather than PG closure. Notably, a recent time-domain spectroscopy study of LSCO \cite{Legros2022}, where estimates of $n_{pl}$ are insensitive to the effects of FS curvature, found a broad crossover in $n_{pl}$, extending over entire strange metal regime, similar to that shown in Fig.~\ref{fig:nH_All}c.

There are several important conclusions that can be drawn from Figure \ref{fig:nH_All}. Firstly, at $p=0.16$, $n_{pl}\approx p$, implying that even at optimal doping, only the doped holes are itinerant. Notably, this value of the normal state carrier density is well below the estimated superfluid density $n_s = m^{\ast} V_\textrm{cell}/2\mu_0 e^2 \lambda_{ab}^2(0)$, where $m^*$ is the effective mass and $\lambda_{ab}(0)$ is the zero temperature in-plane penetration depth. Inserting $m^{\ast}$ = 6$m_e$ \cite{Ramshaw2015} and $\lambda_{ab}(0)$ = 1300~{\AA} \cite{Hossain2012} for optimally doped Y123, we obtain $n_s(0)$ = 0.87, i.e.~around 5 times larger than $n_{pl}$. We discuss this estimate in more detail in the Supplemental Material \cite{SupplementalMaterial} and refer to References~\cite{Leggett65,Varma86,Nomoto13,Walmsley13,Mackenzie1996,Loram1993,Andersen1995,Carrington2007,Rourke2010,Momono1994,Baglo_muSR,Hashimoto2013}. This discrepancy implies that a significant fraction of the SC condensate originates from carriers that do not contribute to the Hall response -- possibly because they are localized, incoherent or particle-hole symmetric. Alternatively, the equivalence of $n_\textrm{H}$ and $n_{pl}$ may be affected by the presence of short-range charge order \cite{Tam2022}.

Arguably the most striking observation, however, is that the $p$ to $1+p$ crossover in Y123 does not terminate at $p^\ast$. When combined with the data from Bi2201 \cite{Putzke2021}, Tl2201 \cite{Putzke2021} and LSCO \cite{Legros2022}, the crossover appears to evolve smoothly across $p^\ast$ with no change of slope. While deviations from this evolution in $n_{\rm H}(p)$ have been observed at low-$T$ in Bi2201 \cite{Balakirev2003}, Y123 \cite{LeBoeuf2007} and LSCO \cite{Balakirev2009} around optimal doping (or below), these can be associated with the onset of charge order within the PG regime. The smooth evolution of $n_{\rm H}(p)$ across $p^\ast$ contrasts markedly with the Hall response in other materials tuned across their respective QCP where sharp changes are invariably observed, even if the character of these changes is different for each material. In electron doped Ba(Fe$_{1-x}$Co$_x$)$_2$As$_2$, for example, the low-$T$ Hall coefficient is found to grow rapidly below the QCP, presumably due to Fermi surface reconstruction induced by the spin density wave order below $x_c = 0.06$ \cite{Liu2010}. In isovalently substituted BaFe$_2$(As$_{1-x}$P$_x$)$_2$ \cite{Hayes2021} and FeSe$_{1-x}$S$_x$ \cite{Culo2021_FeSe}, the anomalous (strange) component in the Hall resistivity exhibits a sharp maximum at the QCP, while in the heavy-fermion YbRh$_2$Si$_2$, $R_{\rm H}$ exhibits a large drop across the (field-induced) QCP, with the width of the drop becoming increasingly sharper as $T$ is reduced \cite{Friedemann2010}. The complete absence of any noticeable feature in $n_{\rm H}$ (or $R_\textrm{H}$) across $p^\ast$, coupled with the incompleteness of the $p$ to $1+p$ crossover at $p^*$, thus suggests the absence of a QCP at $p^*$ (and, by analogy, for the isostructural iridate Sr$_{2-x}$La$_x$IrO$_4$ \cite{Hsu2024}). This would be consistent with there not being a phase transition at finite temperature at $T^*$. 

Before closing, it is important to stress that our interpretation of the Y123 data is critically dependent on the validity of $n_{pl}=n_{\rm H}(\rho_a/\rho_b)^{-1}$. A variety of (semi-classical) effects are known to apply to Y123 that are not captured by the parallel resistor model from which this relation is derived. In the Supplemental Material \cite{SupplementalMaterial} (see also \cite{Shockley1950,Chambers1952,Ong1991,Carrington2007,repository,Badoux2016}), we show the results of a minimal Boltzmann transport model which takes in account the presence of two CuO$_2$ planes per unit cell, the anisotropy of the planes and chains, plane-chain hybridisation in $k$-space and the finite cyclotron motion $\omega_c\tau>0$ at the $1+p$ end of the transition, where no PG exists. The only ingredients necessary to define this model are the Fermi surface shape and the mean free path $\ell$. Using this minimal model, we find in all cases that indeed the rescaling factor results in an estimate much closer to the underlying $n_{pl}$ than the raw $n_{\rm H}$. Thus, the rescaling factor $n_{pl}=n_{\rm H}(\rho_a/\rho_b)^{-1}$ is found to be set by the dimensionality of the planes and chains, and is robust to a wide range of known semi-classical corrections. 

In summary, we have carried out simultaneous measurements of the low-$T$, high-field resistivity anisotropy and Hall effect in overdoped, detwinned single crystals of Y$_{1-x}$Ca$_{x}$Ba$_2$Cu$_3$O$_{7-\delta}$. Renormalisation of the measured carrier density by the resistivity anisotropy results in a gradual $p$ to $1+p$ crossover in $n_{pl}(p)$ that does not terminate at $p^\ast$. This strongly suggests that the `sharp' $p$ to $1+p$ transition seen in $n_{\rm H}(p)$ cannot be construed as evidence for a conventional QCP at $p^\ast$ or a reconstruction of the Fermi surface. Crucially, the gradual $p$ to $1+p$ crossover for Y123 can now be reconciled with similar behavior found in Bi2201 and Tl2201 \cite{Putzke2021} where the full Fermi volume -- as deduced from Hall measurements -- is not recovered until the end of the SC dome. The carrier density in LSCO also appears to follow a similar trend \cite{Legros2022,Culo2021_cuprates}, though the interpretation of $n_\textrm{H}$ is complicated by the presence of the Lifshitz transition ($p\approx0.21$).

The existence of a universal $p$ to $1+p$ crossover can now be viewed as another manifestation of the extended strange metal regime. In Bi2201 and Tl2201, several groups have pointed out that the \lq missing' charge correlates with $T_c$ and the superfluid density, hinting that superconductivity in overdoped cuprates may be dependent upon or driven by a remaining fraction of Mott localised or incoherent carriers \cite{Culo2021_cuprates, Barisic2022, Tromp2023, Ye2024}. Indeed, the combined correlations between $n_{pl}$, the low-$T$ $T$-linear slope of the resistivity, the $H$-linear slope of the high-field MR, the residual superfluid density, the residual specific heat and $T_c$ establish a surprisingly simple, yet highly unconventional phenomenology in overdoped cuprates. This pervasive phenomenology and its occurrence in a region of the phase diagram with ostensibly the simplest electronic structure may be key to unravelling the mysteries of high-$T_c$ superconductivity in hole-doped cuprates.

\begin{acknowledgments}
We thank W. A. Atkinson for helpful discussions. This research was supported by the European Research Council (ERC) under the European Union’s Horizon 2020 research and innovation programme (Grant Agreement No. 835279-Catch-22), the Engineering and Physical Sciences Research Council (EPSRC grant EP/V02986X/1) and a Leverhulme Trust Early Career Fellowship. This work was also supported by HLD-EMFL, member of the European Magnetic Field Laboratory (EMFL) and by EPSRC (UK) via its membership of the EMFL (grant no. EP/N01085X/1).
\end{acknowledgments}

\end{document}